\documentclass[a4paper,12pt,american,twoside]{article}
\usepackage[T1]{fontenc}
\usepackage[latin1]{inputenc}
\usepackage{latexsym}
\usepackage{amsmath}
\usepackage[in,myheadings]{fullpage}
\usepackage{amssymb}
\usepackage{amsfonts}
\usepackage{times}
\usepackage{theorem}
\usepackage{epsfig}
\usepackage{enumerate}
\usepackage{mathrsfs}
\usepackage{latexsym,amsopn}
\usepackage{amsfonts,amsbsy,amscd,stmaryrd}
\usepackage{paralist}

\numberwithin{equation}{section}
\newcounter{resultcounter}[section]

\newtheorem{theorem}{Theorem }[section]
\newtheorem{prop}[theorem]{Proposition}
\newtheorem{lemma}[theorem]{Lemma}
\newtheorem{definition}[theorem]{Definition}
\newtheorem{corollary}[theorem]{Corollary}
\newtheorem{remark}[theorem]{Remark}

 \def\cB{{\cal B}} \def\cC{{\cal C}}
 \def\cE{{\cal E}} 
 \def\cH{{\cal H}} 
\def\cJ{{\cal J}}  \def\cL{{\cal L}}
  
  \def\cR{{\cal R}}
\def\cS{{\cal S}} \def\cT{{\cal T}} \def\cU{{\cal U}}

\def\R{{\mathbb R}}
\def\N{{\mathbb N}}
\def\Q{{\mathbb Q}}
\def\C{{\mathbb C}}
\def\Z{{\mathbb Z}}

\def\D{{\mathbb D}}

\def\slim{\mathop{\rm s-lim}}

\def\tr{{\rm Tr}}
\def\sp{{\rm sp}}
\def\sr{{\rm sr}}
\def\e{{\rm e}}
\def\i{{\rm i}}

\def\pp{{\rm pp}}
\def\ess{{\rm ess}}

\def\bep{\begin{proposition}}
\def\eep{\end{proposition}}
\def\bet{\begin{theoreme}}
\def\eet{\end{theoreme}}
\def\bel{\begin{lemma}}
\def\eel{\end{lemma}}
\def\ss{{\sigma'\sigma}}

\newcommand{\bra}{\langle} 
\newcommand{\ket}{\rangle}

\newcommand{\ds}{\displaystyle}
\def\ran{{\rm Ran}}
\def\sp{{\rm sp}}
\def\rhoinv{\rho_{\beta^*,\cS}}

\def\one{{\mathchoice {\rm 1\mskip-4mu l} {\rm 1\mskip-4mu l} {\rm 1\mskip-4.5mu l} {\rm 1\mskip-5mu l}}}

\def\proof{\noindent {\bf Proof.}\ \ }
\def\qed{\hfill $\Box$\medskip}

\begin{document}
\title{Mixing properties of the one-atom maser}
\author{Laurent Bruneau\footnote{CNRS-UMR 8088 et D\'epartement de Math\'ematiques, Universit\'e de Cergy-Pontoise, Site Saint-Martin, BP 222, 95302 Cergy-Pontoise, France Email: laurent.bruneau@u-cergy.fr, http://bruneau.u-cergy.fr/}
}
\maketitle
\thispagestyle{empty}
\markboth{L. Bruneau}%
{{Slow mixing in a QED cavity}}

\date

\paragraph{Abstract.} We study the relaxation properties of the quantized electromagnetic field in a cavity under repeated interactions with single two-level atoms, so-called one-atom maser. We improve the ergodic results obtained in \cite{BP} and prove that, whenever the atoms are initially distributed according to the canonical ensemble at temparature $T>0$, all the invariant states are mixing. Under some non-resonance condition this invariant state is known to be thermal equilibirum at some renormalized temperature $T^*$ and we prove that the mixing is then arbitrarily slow, in other words that there is no lower bound on the relaxation speed.


\section{Introduction}

During the last years there has been a growing interest for the rigorous development of the quantum statistical mechanics of open systems. Such a system consists in a confined subsystem $\cS$ in contact with
an environment made of one or several extended subsystems $\cR_1,\ldots$ usually called reservoirs. The study of the dynamics of these open quantum systems is an important topic due to its relevance in the description of several basic physical mechanisms of interest, such as convergence towards a thermodynamical equilibrium state of onset of heat or particle fluxes between reservoirs at different temperatures or chemical potentials for example. At the same time, it is a very active field of present research in mathematical physics. One of the reasons for this is  to be found in the fact that the description of return to equilibrium or onset of stationary states in open quantum systems appeals explicitly to the description in the large time regime of the unitary dynamics of quantum systems and the effective dispersive effects induced by the intrinsic properties of the reservoirs. Besides non trivial modeling aspects, the mathematical analysis still represents a challenge for many physically relevant models. We refer the reader to \cite{AJP} and in particular \cite{AJPP1} for a modern introduction to the subject.

Motivated by several new physical applications as well as by their attractive mathematical structure, a class of open systems has recently become very popular in the literature: repeated interaction (RI) systems. There, the environment consists in a sequence $\cE_1$, $\cE_2, \ldots$ of independent subsystems.  The ``small'' subsystem $\cS$ interacts with $\cE_1$ during the time interval $[0,\tau_1[$, then with $\cE_2$ during the interval $[\tau_1,\tau_1+\tau_2[$, etc...  While $\cS$ interacts with $\cE_m$, the other elements of the sequence evolve freely according to their intrinsic (uncoupled) dynamics. Thus, the evolution of the joint system $\cS+\cE_1+\cdots$ is completely determined by the sequence $\tau_1,\tau_2,\ldots$, the individual dynamics of each $\cE_m$ and the coupled dynamics of each pair $\cS+\cE_m$. 

In the simplest case, all the subsystems are identical, {\it i.e.} each $\cE_m$ is the copy of the same $\cE$, and interact with $\cS$ by means of the same coupling operator $V$ on $\cS + \cE$ for the same duration $\tau$.  The dynamics restricted to the small system is shown to be determined by the map $\cL$ which assigns $\rho_\cS(\tau)$ to $\rho_\cS$, see (\ref{def:rdm}), as the result of the interaction of $\cS$ with one subsystem $\cE$ for the duration $\tau$. Heuristically, from the point of view of the small system, all subsystems interacting in sequence with $\cS$ are equivalent, so that the result of $n\in\N$ repeated interactions amounts to iterating  $n$ times the map $\cL$ on the initial condition $\rho_\cS$. This expresses the Markovian character of repeated interactions in discrete time. As a consequence, spectral methods will be available to perform the analysis of the exact dynamics restricted to states on the Hilbert space $\cH_\cS$ of the small system. 
Such models have been analyzed in \cite{BJM1, WBKM} (see also \cite{BJM2} for a random setting). For a pedagogical introduction to RI systems, we refer the reader {\sl e.g.} to \cite{BJM3}. Let us also note here that when the dimension of $\cH_\cS$ is finite the spectral analysis of the map $\cL$ is, in principle, straightforward. However, in case $\cH_\cS$ is infinite dimensional, as in the present paper, it becomes much more delicate.

The physical situation which is perhaps the most tightly linked to the repeated interaction models is that of the one atom maser \cite{FJM, CDG, MWM, WVHW, WBKM}, and some of its subsequent elaborations  \cite{DRBH, G-al, RH, RBH}.
Here, $\cS$ is the quantized electromagnetic field of a cavity through which a beam of atoms, the $\cE_m$, is shot in such a way that no more than {\sl one atom} is present in the cavity at any time. Such systems play a fundamental role in the experimental and theoretical investigations of basic matter-radiation processes. They are also of practical importance in quantum optics and quantum state engineering \cite{MWM,WVHW,WBKM,RH,VAS}. So-called ``One-Atom Masers'', where the beam is tuned in such a way that at each given moment a single atom is inside a microwave cavity and the interaction time $\tau$ is the same for each atom, have been experimentally realized in laboratories \cite{MWM,WVHW}.

In this paper we continue the mathematical analysis of a specific model of RI system describing the one-atom maser experiment mentioned above and initiated in \cite{BP} (the model is described in Section \ref{sec:model}). The first natural question is that of thermal relaxation (a question which has been extensively studied when a small system $\cS$ with a \emph{finite} dimensional Hilbert space is coupled to an ideal quantum gas, see e.g. \cite{JP1,BFS,DJ1,FM}): is it possible to thermalize a mode of a QED cavity by means of $2$-level atoms if the latter are initially at thermal equilibrium? It is proven in \cite{BP} that the answer is positive but the relaxation was proven only in a mean ergodic sense. Numerical simulations however indicated that such an ergodic average was not needed, in other words the thermal equilibrium state of the cavity field (at a temperature dictated by the one of the $2$-level atoms) was not only ergodic but mixing. These simulations also showed that the relaxation would be slow due to the presence of infinitely many meta-stable states, see Section \ref{ssec:Lbetanot}. The purpose of the present article is to prove these two facts.  As in \cite{BP}, we would like to emphasize that in our situation the Hilbert space of the small system $\cS$ is \emph{not} finite dimensional. Moreover, we do not make use of any perturbation theory, {\sl i.e.,} our results do not restrict to small coupling constants.

The paper is organized as follows. The precise description of the model is given in Section \ref{sec:model}. In Section \ref{sec:mixing} we recall some of the important features of the model and state our main results (Theorems \ref{thm:strongmixing} and \ref{thm:slowmixing}). The proof of Theorem \ref{thm:strongmixing} will be found in Section \ref{sec:proof1} and the one of Theorem \ref{thm:slowmixing} in Section \ref{sec:proof2}.

{\bf Acknowledgements.} The author is grateful to V. Georgescu for fruitful discussions and to C. Pellegrini for drawing his attention to reference \cite{GvH}. This work was partially supported by the Agence Nationale de la Recherche, grant ANR-09-BLAN-0098-01.


\section{Description of the model}\label{sec:model}

\subsection{The Jaynes-Cummings Hamiltonian}

The atoms of the beam are prepared in a stationary mixture of two states with energies $E_0<E_1$ and we assume the cavity to be nearly resonant with the transitions between these two states. Neglecting the non-resonant modes of the cavity, we can describe its quantized electromagnetic field by a single harmonic oscillator of frequency $\omega\simeq\omega_0\equiv E_1-E_0$.

The Hilbert space of the cavity field is $\cH_\cS\equiv\ell^2(\mathbb N)=\Gamma_+(\C)$, the Bosonic Fock space over $\C$. Its Hamiltonian is
$$
H_\cS\equiv\omega N\equiv\omega a^* a,
$$
where $a^*$, $a$ are the creation/annihilation operators on $\cH_\cS$ satisfying the commutation relation $[a,a^*]=\one$. Normal states of $\cS$ are density matrices, positive trace class operators $\rho$ on $\cH_\cS$ with $\tr\rho=1$. As in \cite{BP}, these are the only states we shall consider on $\cS$.  Therefore, in the following, ``state'' always means ``normal state'' or equivalently ``density matrix''. Moreover, we will say that a state is diagonal if it is represented by a diagonal matrix in the eigenbasis of $H_\cS$.

The Hilbert space for a single atom is $\cH_\cE := \C^2$ which, for notational convenience, we identify with $\Gamma_-(\C)$, the Fermionic Fock space over $\C$. Without loss of generality we set $E_0=0$. The Hamiltonian of a single atom is thus
$$
H_\cE:= \omega_0b^*b,
$$
where $b^*$, $b$ denote the creation/annihilation operators on $\cH_\cE$. Stationary states of the atom can be parametrized by the inverse temperature $\beta\in\R$ and are given by the density matrices $\rho_{\beta,\cE}:={\e^{-\beta H_\cE}}/{\tr \ \e^{-\beta H_\cE}}$.

In the dipole approximation, an atom interacts with the the cavity field through its electric dipole moment. The full dipole coupling is given by $(\lambda/2)(a+a^*)\otimes(b+b^*)$, acting on $\cH_\cS\otimes \cH_\cE$, where $\lambda\in\R$ is a coupling constant. Neglecting the counter rotating term $a\otimes b+a^*\otimes b^*$ in this coupling (this is the so called {\sl rotating wave approximation}) leads to the well known Jaynes-Cummings Hamiltonian
\begin{equation*}
H\equiv H_\cS\otimes \one_\cE+\one_\cS\otimes H_\cE+\lambda V,
\qquad
V\equiv\frac{1}{2}(a^*\otimes b+a\otimes b^*),
\label{eq:interham}
\end{equation*}
for the coupled system $\cS+\cE$ (see {\sl e.g.,} \cite{Ba,CDG,Du}). The operator $H$ has a distinguished property which allows for its
explicit diagonalisation: it commutes with the total number operator 
\begin{equation}
N_{\rm tot}:= a^*a+b^*b.
\label{Mdef}
\end{equation}


\subsection{The one-atom maser model}

Given an interaction time $\tau>0$, the system $\cS$ successively interacts with different copies of the system $\cE$, each interaction having a duration $\tau$. The issue is to understand the asymptotic behavior of the system $\cS$ when the number of such interactions tends to $+\infty$ (which is equivalent to time $t$ going to $+\infty$). The Hilbert space describing the entire system $\cS+\cC$ is then
$$
\cH:=\cH_\cS\otimes\cH_\cC,\qquad \cH_\cC :=\bigotimes_{n\geq 1} \cH_{\cE_n},
$$
where $\cH_{\cE_n}$ are identical copies of $\cH_\cE$.  During the time interval $[(n-1)\tau,n\tau)$, the system $\cS$ interacts only with the $n$-th element of the chain. The evolution is thus described by the Hamiltonian $H_n$ which acts as $H$ on $\cH_\cS\otimes\cH_{\cE_n}$ and as the identity on the other factors $\cH_{\cE_k}$.

\begin{remark} A priori we should also include the free evolution of the non-interacting elements of $\cC$. However, since we shall take the various
elements of $\cC$ to be initially in thermal equilibrium, this free evolution will not play any role.
\end{remark}

Given any initial  state $\rho$ on $\cS$ and assuming that all the atoms are in the stationary state $\rho_{\beta,\cE}$, the state of the total repeated interaction system after $n$ interactions is thus given by
$$
\e^{-i\tau H_n}\cdots\e^{-i\tau H_1} \left( \rho\otimes \bigotimes_{k\geq 1} \rho_{\beta,\cE}\right) \e^{i\tau H_1}\cdots\e^{i\tau H_n}.
$$
To obtain the state $\rho_n$ of the system $\cS$ after these $n$ interactions we take the partial trace over the chain $\cC$, {\sl i.e.,}
\begin{equation*}
\label{eq:systevol}
\rho_n=\tr_{\cH_\cC}\left[ \e^{-i\tau H_n}\cdots\e^{-i\tau H_1} \left(\rho\otimes \bigotimes_{k\geq 1} \rho_{\beta,\cE} \right) 
\e^{i\tau H_1}\cdots\e^{i\tau H_n} \right].
\end{equation*}
It is easy to make sense of this formal expression (we deal here with countable tensor products). Indeed, at time $n\tau$ only the $n$ first elements of the chain have played a role so that we can replace $\bigotimes_{k\geq 1} \rho_{\beta,\cE}$ by $\rho_{\beta,\cE}^{(n)}:=\bigotimes_{k=1}^n \rho_{\beta,\cE}$ and the partial trace over the chain by the partial trace over the finite tensor product $\cH_{\cC}^{(n)}\equiv\bigotimes_{k=1}^n \cH_{\cE_k}$.

The very particular structure of the repeated interaction systems allows to rewrite $\rho(n)$ in a much more convenient way. The two main characteristics of these systems are:
\begin{compactenum}
\item The various subsystems of the environment do not interact directly (only via $\cS$),
\item The system $\cS$ interacts only once with each subsystem $\cE_n$, and with only one at a time.
\end{compactenum}
It is therefore easy to see that the evolution of the system $\cS$ is Markovian: the state $\rho_n$ only depends on the state $\rho_{n-1}$
and the $n$-th interaction (see \cite{AJ1,BJM1,BP}). More precisely, we have
$$
\rho_n=\cL_\beta(\rho_{n-1}),
$$
where
\begin{equation}
\label{def:rdm}
\cL_\beta(\rho):= \tr_{\cH_\cE}\left[ \e^{-i\tau H} (\rho\otimes \rho_{\beta,\cE})\, \e^{i\tau H} \right].
\end{equation}

\begin{definition} The map $\cL_\beta$ defined on the set $\cJ_1(\cH_\cS)$ of trace class operators on $\cH_\cS$ by (\ref{def:rdm}) is called the reduced dynamics.  The state of $\cS$ evolves according to the discrete semigroup $\{\cL_\beta^n\,|\,n\in\N\}$ generated by this map:
$$
\rho_n=\cL_\beta^n(\rho).
$$
In particular, a state $\rho$ is invariant iff $\cL_\beta(\rho)=\rho$.
\end{definition}

Note that $\cL_\beta$ is clearly a contraction. To understand the asymptotic behavior of $\rho_n$, one has to understand its spectral properties and in particular its peripheral spectrum $\sp(\cL_\beta)\cap S^1$. 

\begin{remark}\label{rem:uncoupledspectrum} When the atom-field coupling is turned off, the reduced dynamics is nothing but the free evolution of $\cS$, {\sl i.e.} $\cL_\beta(\rho)=\e^{-i\tau H_\cS}\rho\, \e^{i\tau H_\cS}$. Thus, for $\lambda=0$, the spectrum of $\cL_\beta$ is pure point
$$
\sp(\cL_\beta)=\sp_\pp(\cL_\beta)=\overline{\{\e^{\i\tau\omega d}\,|\,d\in\Z\}}.
$$
This spectrum is finite if $\tau\omega\in2\pi\Q$ and densely fills the unit circle in the opposite case. In both cases, all the eigenvalues, and in particular $1$, are infinitely degenerate: one has, for any $n$,
$$
\cL_\beta(|n\ket\bra n+d|) =\e^{\i\tau\omega d} |n\ket\bra n+d|.
$$
This explains why perturbation theory in $\lambda$ fails for this model. As we shall see, this spectrum will actually survive after turning on the coupling, i.e. for any $\lambda$ one has $\overline{\{\e^{\i\tau\omega d}\,|\,d\in\Z\}} \subset \sp(\cL_\beta)$, even though only $1$ stays as an eigenvalue.
\end{remark}

\begin{remark} A similar model has also recently been studied in \cite{NVZ}, with a coupling operator of the form $V=(a+a^*)\otimes b^*b$. The latter has the advantage to leave invariant the state of the atom and therefore leads to more tractable computations. 
\end{remark}


\section{Mixing properties of $\cL_\beta$}\label{sec:mixing}


\subsection{Rabi resonances}\label{ssec:resonances}

An essential feature of the dynamics generated by the Jaynes-Cummings Hamiltonian are Rabi oscillations. In the presence of $n$ photons, the probability for the atom to make a transition from its ground state to its excited state is a periodic function of time. The circular frequency of this oscillation is given by $\nu_n:=\sqrt{\lambda^2 n+(\omega_0-\omega)^2}$, a fact easily derived from the propagator formula (\ref{Propa}) below. These oscillations are at the origin of what was called a \emph{Rabi resonance} in \cite{BP}. Such a resonance occurs when the interaction time $\tau$ is an integer multiple of the period of a Rabi oscillation, i.e. $\tau=k\frac{2\pi}{\nu_n}$ for some $k\in\N$. In terms of the dimensionless detuning parameter and coupling constant
$$
\eta:= \left(\frac{\Delta\tau}{2\pi}\right)^2,\qquad \xi:=\left(\frac{\lambda\tau}{2\pi}\right)^2,
$$
a positive integer $n$ is a Rabi resonance if
\begin{equation}
\xi n+\eta=k^2,
\label{lock}
\end{equation}
for some positive integer $k$. Depending on the arithmetic properties of $\eta$ and $\xi$ one easily proves (\cite{BP}, Lemma 3.2) that the system has either no, one or infinitely many Rabi resonances. Accordingly, the system is called non-resonant, simply resonant or fully resonant. If $R(\eta,\xi)$ denotes the set of Rabi resonances, the Hilbert space $\cH_\cS$ has a decomposition
\begin{equation*}
\cH_S=\bigoplus_{k=1}^r\cH_\cS^{(k)},
\label{RabiDecomp}
\end{equation*}
where $r-1$ is the number of Rabi resonances, $\cH_\cS^{(k)}\equiv\ell^2(I_k)$ and $\{I_k\,|\,k=1,\ldots, r\}$ is the partition of $\N$ induced by the resonances. $\cH_\cS^{(k)}$ is called the $k$-th Rabi sector, and $P_k$ denotes the corresponding orthogonal projection.


\subsection{Ergodicity and mixing}\label{ssec:ergodicity}

In \cite{BP} we investigated the ergodic properties of the map $\cL_\beta$ and of its invariant states. For any density matrix $\rho$, we denote the
orthogonal projection on the closure of $\ran\,\rho$ by $s(\rho)$, the support of $\rho$. We also write $\mu\ll\rho$ whenever $s(\mu)\le
s(\rho)$. A state $\rho$ is called ergodic, respectively mixing, for the semigroup generated by $\cL_\beta$ whenever
\begin{equation}
\lim_{N\to\infty}\frac{1}{N}\,\sum_{n=1}^N \tr\left(\cL_\beta^n(\mu)\, A\right)=\tr(\rho\, A),
\label{ergodef}
\end{equation}
respectively
\begin{equation}
\lim_{n\to\infty}\tr\left(\cL_\beta^n(\mu)\, A\right)=\tr(\rho\, A),
\label{mixdef}
\end{equation}
holds for all states $\mu\ll\rho$ and all $A\in\cB(\cH_\cS)$. $\rho$ is exponentially mixing if the convergence in (\ref{mixdef}) is exponential, {\sl i.e.,} if
$$
\left|\tr\left(\cL_\beta^n(\mu)\, A\right)-\tr(\rho\, A)\right|\le C_{A,\mu}\,\e^{-\alpha n},
$$
for some constant $C_{A,\mu}$ which may depend on $A$ and $\mu$ and some $\alpha >0$ independent of $A$ and $\mu$. A mixing state is ergodic and an ergodic state is clearly invariant. 

For $\beta\in\R$ we set $\beta^*:=\beta\omega_0/\omega$ and to each Rabi sector $\cH_\cS^{(k)}$ we associate the state
$$
\rho^{(k)}_{\beta^*,\cS}:= \frac{\e^{-\beta^* H_\cS}P_k}{\tr \ \e^{-\beta^* H_\cS}P_k} =\frac{\e^{-\beta\omega_0 N}P_k}{\tr \ \e^{-\beta\omega_0N}P_k}.
$$

The following theorem is (part of) the main result of \cite{BP}. It relies on the analysis of the peripheral eigenvalues of $\cL_\beta$.
\begin{theorem}\label{thm:BP} [BP] 1. If the system is non-resonant then $\cL_\beta$ has no invariant state for $\beta\leq0$ and the unique ergodic state
$\ds \rho_{\beta^*,\cS}=\frac{\e^{-\beta^* H_\cS}}{\tr \ \e^{-\beta^* H_\cS}}$ for $\beta>0$. In the latter case any initial state relaxes in the mean to the thermal equilibrium state at inverse temperature $\beta^*$.

2. If the system is simply resonant, then $\cL_\beta$ has the unique ergodic state $\rho^{(1)}_{\beta^*,\cS}$ if $\beta\le0$ 
and two ergodic states $\rho^{(1)}_{\beta^*,\cS}$,  $\rho^{(2)}_{\beta^*,\cS}$ if $\beta>0$.

3. If the system is fully resonant then for any $\beta\in\R$, $\cL_\beta$ has infinitely many ergodic states  $\rho^{(k)}_{\beta^*,\cS}$, $k=1,2,\ldots$.

4. If the sector $\cH_\cS^{(k)}$ is finite dimensional the state $\rho^{(k)}_{\beta^*,\cS}$ is exponentially mixing.
\end{theorem}

Numerical experiments support the conjecture that actually even in infinite dimensional sectors the ergodic states are mixing. Moreoever the map $\cL_\beta$ has an infinite number of metastable states in the non-resonant and simply resonant cases (see Section 4.5 in \cite{BP} and Section \ref{ssec:essentialradius}). As a result one expects slow, {\sl i.e.} non-exponential, mixing. The purpose of the present paper is to prove these two facts: 
\begin{compactenum}
 \item all the ergodic states are mixing,
 \item if the sector $\cH_\cS^{(k)}$ is infinite dimensional then $\rho^{(k)}_{\beta^*,\cS}$ is slowly mixing.
\end{compactenum}

From now on, we will only consider the non-resonant case. In the simply resonant case, to prove that the invariant state $\rho_{\beta^*,\cS}^{(2)}$ is  (slowly) mixing, it suffices to consider the restriction of $\cL_\beta$ to the second Rabi sector, i.e. to $\cJ_1\left(\cH_\cS^{(2)}\right)$, and we then come back to a non-resonant situation. Our main result is
\begin{theorem}\label{thm:strongmixing} Suppose the system is non-resonant and let $\beta>0$. Then for any initial state $\rho$ one has
\begin{equation}\label{eq:strongmixing}
\lim_{n\to\infty} \left\|\cL_\beta^n(\rho)-\rhoinv\right\|_1=0.
\end{equation}
In particular, the unique invariant state $\rhoinv$ of $\cL_\beta$ is mixing. In other words, any initial state relaxes to the thermal equilibrium state at inverse temperature $\beta^*$.
\end{theorem}


\subsection{Arbitrarily slow mixing}\label{ssec:slowmixing}

The next question concerning the mixing properties of $\cL_\beta$ is that of the speed of convergence in (\ref{eq:strongmixing}). As mentioned in the previous section, there is numerical evidence that this convergence is slow which is due to the presence of an infinite number of metastable states with arbitrarily large life-time. We shall give partial information in this direction, showing that not only the mixing is not exponential but that there is no lower bound on the speed of convergence.

In order to state our result about the slowness of convergence in (\ref{eq:strongmixing}), we introduce the notion of arbitrarily slow convergence. Following \cite{BGM}, if $T$ is an operator such that $T^n\to T_\infty$ in the strong sense, we shall say that it satisfies
\begin{itemize}
\item[{\bf (ASC)}] \  If for any sequence $(\epsilon_n)_n$ of positive numbers such that $\lim \epsilon_n=0$ there exists a vector $x$ and a linear form $\varphi$ such that for $n$ large enough
\begin{equation}\label{eq:ascw}
|\bra \varphi, T^nx-T_\infty x \ket |\geq \epsilon_n.
\end{equation}
\end{itemize}
(ASC) stands for arbitrarily slow convergence and corresponds to condition (ASC3) in \cite{BGM}.

\begin{theorem}\label{thm:slowmixing} Suppose the system is non-resonant and let $\beta>0$. Then $\cL_\beta$ satisfies (ASC). More precisely, for any sequence $(\epsilon_n)_n$ of positive numbers such that $\lim \epsilon_n=0$, there exists an initial state $\rho$, an observable $A$, $C>0$ and $n_0\in\N$ such that, 
\begin{equation}\label{eq:slowmixing}
\left|\tr\left( \cL_\beta^n(\rho)\, A\right) - \tr \left(\rhoinv \, A\right) \right| \geq C\epsilon_n, \qquad \forall n\geq n_0.
\end{equation}
\end{theorem}

The above theorem precisely says that there is no lower bound on the convergence speed in Theorem \ref{thm:strongmixing} and in particular that the mixing is not exponential. It is still an open question to get an upper bound on this convergence speed.

\begin{remark}\label{rem:slowdecoherence} As mentioned in \cite{BP}, Theorem \ref{thm:strongmixing} implies in particular decoherence in the energy eigenbasis of the cavity field. We will see on Section \ref{ssec:slowmixingproof} that this decoherence too can be arbitrarily slow. 
\end{remark}


\section{Proof of Theorem \ref{thm:strongmixing}}\label{sec:proof1}


\subsection{Strategy of the proof}\label{ssec:strategy}

The ergodic properties obtained in \cite{BP} rely on the analysis of the peripheral eigenvalues of the operator $\cL_\beta$. The main obstacle to the proof of mixing is the lack of information concerning the peripheral spectrum of the operator $\cL_\beta$ (only information about peripheral \emph{eigenvalues} were obtained). 

To remedy it, the idea is to consider the dual map $(\cL_\beta)^*$ on $\cB(\cH_\cS)$, actually the dual map in the interaction picture (see \ref{def:interactionpicture}), and, following \cite{CF,GvH}, consider its representation $L_\beta$ in the following embedding of $\cB(\cH_\cS)$ into the space $\cJ_2(\cH_\cS)$ of Hilbert-Schmidt operators:
$$
\Phi: \cB(\cH_\cS) \ni X \mapsto \rho_{\beta^*,\cS}^{1/4}X\rho_{\beta^*,\cS}^{1/4}  \in \cJ_2(\cH_\cS),
$$
i.e. $L_\beta$ is such that $L_\beta \left(\Phi(X)\right) = \Phi\left(\cL_\beta^*(X) \right)$. 

We prove in Section \ref{ssec:Lbetadef} that $L_\beta$ extends to a contraction on $\cJ_2(\cH_\cS)$ and using the gauge invariance, w.r.t. to the gauge group $\e^{-\i\theta N}\cdot \e^{\i\theta N}$, we show in Section \ref{ssec:Lbetagaugeinv} that it leaves invariant the subspaces
$$
\cJ_2^{(d)}(\cH_\cS):= \{X \in \cJ_2(\cH_\cS)\,|\,\e^{-\i\theta N} X \e^{\i\theta N} =\e^{\i\theta d}X  \ \text{for all}\ \theta\in\R\},
$$
and analyze separately the spectrum of $L_\beta$ on each of these subspaces. The restriction $L_\beta^{(0)}$ of $L_\beta$ on $\cJ_2^{(0)}(\cH_\cS)$ is studied in Section \ref{ssec:Lbetanot}. We show that it is self-adjoint and satisfies $-1<\alpha\leq L_\beta^{(0)}\leq 1$. Then for any $d$ we prove in Section \ref{ssec:Lbetad} that it can be written as a compact perturbation of $L_\beta^{(0)}$ (up to considering both operators as acting on $\ell^2(\N)$). As a consequence, except maybe at $1$, the peripheral spectrum of $L_\beta^{(d)}$ consists only in peripheral eigenvalues. We then rule out such peripheral eigenvalues as in \cite{BP} using the following Perron-Frobenius type theorem due to Schrader (\cite{sch}, Theorem 4.1)
\begin{theorem}\label{thm:schrader}Let $\phi$ be a $2$-positive map on $\cJ_2(\cH)$ such that $\sr(\phi)=\|\phi\|$. If $\lambda$ is a peripheral eigenvalue of $\phi$ with eigenvector $X$, \sl i.e. $\phi(X)=\lambda X$, $X\not=0$, $|\lambda|=\sr(\phi)$, then $|X|=\sqrt{X^*X}$ is an eigenvector of $\phi$ to the eigenvalue $r(\phi)$: $\phi(|X|)=r(\phi)|X|$.
\end{theorem}

We use these spectral information to derive mixing properties of $L_\beta$ in Section \ref{ssec:Lbetamixing} and finally deduce similar mixing properties for the operator $\cL_\beta$ in Section \ref{sec:proofthm1}. Our main tool to go from the spectrum of $L_\beta$ to its mixing properties is the following theorem due to Badea, Grivaux and M\"uller \cite{BGM}. 

\begin{theorem}\label{thm:bgm1} Let $Z$ be a Banach space and $T\in\cB(Z)$ a power bounded mean ergodic operator with spectrum $\sp(T)$ included in $\D\cup \{1\}$, where $\D$ denotes the open unit disk in the complex plane. Then the sequence of iterates $T^n$ is strongly convergent to the projection onto $\ker(T-\one)$ along $\overline{\ran(T-\one)}$.
\end{theorem}

This theorem is actually an almost immediate consequence of a celebrated theorem due to Katznelson and Tzafiri \cite{KT} which asserts that if $T$ is a contraction on a Banach space then $\ds \lim_{n\to\infty} \|T^n-T^{n+1}\|=0$ if and only if $\sp(T)\cap S^1 \subset \{1\}$.


\subsection{Gauge invariance and Kraus representation of $\cL_\beta$}

It follows from its definition, see (\ref{def:rdm}), that the map $\cL_\beta$ is a trace preserving completely positive map on $\cJ_1(\cB_\cS)$.  

Denote by $|-\ket$ and $|+\ket$ the ground state and the excited state of the atom $\cE$. This orthonormal basis of $\cH_\cE$ allows us to
identify $\cH=\cH_\cS\otimes\cH_\cE$ with $\cH_S\oplus\cH_S$. Using the fact that $H$ commutes with the total number operator $N_{\rm tot}$ (recall
(\ref{Mdef})), an elementary calculation shows that, in this representation, the unitary group $\e^{-\i\tau H}$ is given by
\begin{equation}
\e^{-\i\tau H}=
\left(
\begin{matrix} \ds\e^{-\i(\tau\omega N+\pi\eta^{1/2})}\,C(N)&
\ds-\i\e^{-\i(\tau\omega N+\pi\eta^{1/2})}S(N)\,a^*\\[13pt]
\ds-\i\e^{-\i(\tau\omega(N+1)+\pi\eta^{1/2})}S(N+1)\,a & 
\e^{-\i(\tau\omega(N+1)+\pi\eta^{1/2})}\,C(N+1)^*
\end{matrix}
\right),
\label{Propa}
\end{equation}
where
\begin{equation}\label{def:CSfunctions}
C(N):=\cos(\pi\sqrt{\xi N+\eta})+\i\eta^{1/2}\,\frac{\sin(\pi\sqrt{\xi N+\eta})}{\sqrt{\xi N+\eta}},\quad
S(N):=\xi^{1/2}\,\frac{\sin(\pi\sqrt{\xi N+\eta})}{\sqrt{\xi N+\eta}},
\end{equation}
with the convention $\sin(0)/0=1$ to avoid any ambiguity in the case $\eta=0$. Let $w_\beta(\sigma)\equiv\bra\sigma|\rho_\cE^\beta|\sigma\ket
=(1+\e^{\sigma\beta\omega_0})^{-1}$ denote the Gibbs distribution of the atoms. The defining identity (\ref{def:rdm}) yields
\begin{equation}
\cL_\beta(\rho)=\sum_{\sigma,\sigma'}\bra\sigma'|\e^{-\i\tau H}|\sigma\ket w_\beta(\sigma)\rho \bra\sigma|\e^{\i\tau H}|\sigma'\ket=
\sum_{\sigma,\sigma'} V_\ss \rho V_\ss^*,
\label{eq:meanstate}
\end{equation}
where the operators $V_\ss$ are given by 
\begin{equation}\label{eq:vsigma}
\begin{array}{ll}
\ds V_{--}=w_\beta(-)^{1/2}\,\e^{-\i\tau\omega N}\,C(N),& 
\ds V_{-+}=w_\beta(+)^{1/2}\,\e^{-\i\tau\omega N}\,S(N)\,a^*, \\[16pt]
\ds V_{+-}=w_\beta(-)^{1/2}\,\e^{-\i\tau\omega N}\,S(N+1)\,a,& 
\ds V_{++}=w_\beta(+)^{1/2}\,\e^{-\i\tau\omega N}\,C(N+1)^*.
\end{array}
\end{equation}
The above formulas give an explicit Kraus representation of the CP map $\cL_\beta$, see e.g. \cite{Kr,sch,st}. Using the facts that $[H,N_{\rm tot}]=[H_\cE,\rho_{\beta,\cE}]=0$, one also easily shows from the definition (\ref{def:rdm}) that
\begin{equation}\label{eq:gauge}
\cL_\beta(\e^{-\i\theta N}X\e^{\i\theta N}) =\e^{-\i\theta N}\cL_\beta(X)\e^{\i\theta N},
\end{equation}
holds for any $X\in\cJ_1(\cH_\cS)$ and $\theta\in\R$.


\subsection{ The $\cJ_2$ embedding: the operator $L_\beta$}\label{ssec:Lbetadef}

As we mentioned, we shall not directly study the peripheral spectrum of $\cL_\beta$ but the one of a closely related operator $L_\beta$ (the representation in $\cJ_2(\cH_\cS)$ of the adjoint of $\cL_\beta$ in the interaction picture) which we now describe more precisely.

Introducing the non-interacting evolution operator
\begin{equation}\label{def:freeevol}
\cU(\rho)=\e^{-i\tau H_\cS} \rho \, \e^{i\tau H_\cS} = \e^{-i\omega\tau N} \rho \, \e^{i\omega\tau N},
\end{equation}
we define the reduced dynamics in the interaction picture as 
\begin{equation}\label{def:interactionpicture}
\tilde \cL_\beta := \cL_\beta \circ \cU^{-1}.
\end{equation}
Using (\ref{eq:gauge}) we get
\begin{equation}\label{eq:dynamicsplitting}
\cL_\beta^n = \tilde \cL_\beta^n \circ \cU^{-n} = \cU^{-n}\circ \tilde\cL_\beta^n
\end{equation}
for any $n$.

Let $\tilde\cL_\beta^*$ denote the adjoint of $\tilde\cL_\beta$ w.r.t. to the duality $\bra A|\rho\ket = \tr(A\rho)$. The map $\tilde\cL_\beta^*$ acts on $\cB(\cH_\cS)$, i.e. on observables. The map $\tilde \cL_\beta^*$ is also a CP map whose Kraus representation is given by
\begin{equation}
\tilde\cL_\beta^*(A)= \sum_{\sigma,\sigma'} \tilde V_\ss^* A \tilde V_\ss,
\label{eq:meanstate2}
\end{equation}
where $\tilde V_\ss=\e^{\i\omega\tau N} V_\ss$ for any $\sigma,\sigma'$.

Consider now the following embedding of $\cB(\cH_\cS)$ into $\cJ_2(\cH_\cS)$:
\begin{equation}\label{def:embedding}
\Phi: \cB(\cH_\cS) \ni A \mapsto \rho_{\beta^*,\cS}^{1/4}A\rho_{\beta^*,\cS}^{1/4} \in \cJ_2(\cH_\cS).
\end{equation}
Since $\rhoinv>0$, $\Phi$ is injective and on $\ran(\Phi)$ we define $L_\beta$ by
\begin{equation}\label{def:Lbeta}
L_\beta \left( \Phi(A)\right) : = \Phi \circ \tilde\cL_\beta^*(A). 
\end{equation}

In the sequel we shall simply write $\cJ_2$ for $\cJ_2(\cH_\cS)$.
\begin{lemma}\label{lem:Lbetaconraction} The operator $L_\beta$ extends to a contraction on $\cJ_2$. 
\end{lemma}

\proof  Let $X=\Phi(A)\in \ran(\Phi)$ so that $L_\beta(X)=\Phi\circ \tilde\cL_\beta^*(A)$. For any $Y\in \cJ_2$ we have
\begin{eqnarray*}
\lefteqn{ \bra Y, L_\beta(X) \ket_{\cJ_2}}\\
 & = & \tr_\cS\left[ Y^* L_\beta(X)\right] \\
 & = & \tr_\cS \left[ Y^* \rhoinv^{1/4} \tilde \cL_\beta^*(A) \rhoinv^{1/4} \right] \\
 & = & \tr_\cS \left[  \cL_\beta(\e^{i\tau H_\cS}\rhoinv^{1/4} Y^*\rhoinv^{1/4} \e^{-i\tau H_\cS}) A \right] \\
 & = & \tr_{\cS\otimes \cE} \left[ \e^{-i\tau H} \left( \rhoinv^{1/4} \e^{i\tau H_\cS}Y^*\e^{-i\tau H_\cS}\rhoinv^{1/4} \otimes \rho_{\beta,\cE} \right) \e^{i\tau H}  \left(A \otimes \one_\cE \right) \right] \\
 & = & \tr_{\cS\otimes \cE} \left[ \e^{-i\tau H} \left( \rhoinv^{1/4} \otimes \rho_{\beta,\cE}^{1/4}\right) \left(\e^{i\tau H_\cS}Y^*\e^{-i\tau H_\cS} \otimes \rho_{\beta,\cE}^{1/2} \right) \left( \rhoinv^{1/4} \otimes \rho_{\beta,\cE}^{1/4}\right) \e^{i\tau H} \left(A \otimes \one_\cE \right) \right] \\
 & = & \tr_{\cS\otimes \cE} \left[\left( \rhoinv^{1/4} \otimes \rho_{\beta,\cE}^{1/4}\right)  \e^{-i\tau H}  \left(\e^{i\tau H_\cS}Y^* \e^{-i\tau H_\cS}\otimes \rho_{\beta,\cE}^{1/2} \right) \e^{i\tau H} \left(\rhoinv^{1/4} A \otimes \rho_{\beta,\cE}^{1/4}\right) \right] \\
 & = & \tr_{\cS\otimes \cE} \left[ \e^{-i\tau H}  \left(\e^{i\tau H_\cS}Y^* \e^{-i\tau H_\cS}\otimes \rho_{\beta,\cE}^{1/2} \right) \e^{i\tau H}  \left(X  \otimes \rho_{\beta,\cE}^{1/2}\right) \right], 
\end{eqnarray*}
where we have used (\ref{def:embedding})-(\ref{def:Lbeta}) in the second equality, the cyclicity of the trace and (\ref{def:freeevol})-(\ref{def:interactionpicture}) in the third one, (\ref{def:rdm}) in the fourth one, the fact that $\ds \rhoinv\otimes \rho_{\beta,\cE} = \frac{\e^{-\beta \omega_0 N_{\rm tot}}}{\tr \left(\e^{-\beta \omega_0 N_{\rm tot}} \right)}$ commutes with $\e^{it H}$ in the fifth one, and the cyclicity of the trace again in the last one.

Therefore, since $\e^{i\tau H_\cS}$ and $\e^{i\tau H}$ are unitary, we have for any $Y\in \cJ_2$
\begin{eqnarray*}
\left|\bra Y, L_\beta(X) \ket_{\cJ_2}\right| & \leq & \left\|\e^{i\tau H_\cS}Y \e^{-i\tau H_\cS}\otimes \rho_{\beta,\cE}^{1/2} \right\|_{\cJ_2(\cH_\cS\otimes \cH_\cE)} \times \left\|X  \otimes \rho_{\beta,\cE}^{1/2}\right\|_{\cJ_2(\cH_\cS\otimes \cH_\cE)}\\
 & = & \|Y\|_{\cJ_2(\cH_\cS)} \times \|X\|_{\cJ_2(\cH_\cS)} \times \|\rho_{\beta,\cE}^{1/2}\|_{\cJ_2(\cH_\cE)}^2 \\
 & = & \|Y\|_{\cJ_2(\cH_\cS)} \times \|X\|_{\cJ_2(\cH_\cS)},
\end{eqnarray*}
and hence $\|L_\beta(X)\|_{\cJ_2(\cH_\cS)} \leq \|X\|_{\cJ_2(\cH_\cS)}$. The operator $L_\beta$ defines a contraction on $\ran(\Phi)$ and thus extends to a contraction on $\cJ_2$.
\qed

Note that $\ran(\Phi)$ is dense in $\cJ_2$ (it contains all finite rank operators since $\rhoinv$ is faithful) so this extension is actually unique.

It now easily follows from (\ref{eq:vsigma}), (\ref{eq:meanstate2}) and (\ref{def:Lbeta}) that
\begin{equation}
L_\beta(X)= \sum_{\sigma,\sigma'} \hat V_\ss^* X \hat V_\ss, \qquad \forall X\in \cJ_2,
\label{eq:meanstate3}
\end{equation}
where the operators $\hat V_\ss$ are given by 
\begin{equation}\label{eq:vsigmahat}
\begin{array}{ll}
\ds \hat V_{--}=\frac{1}{\sqrt{Z_\beta}}\,C(N),& 
\ds \hat V_{-+}=\frac{\e^{-\beta\omega_0/4}}{\sqrt{Z_\beta}}\,S(N)\,a^*, \\[16pt]
\ds \hat V_{+-}=\frac{\e^{-\beta\omega_0/4}}{\sqrt{Z_\beta}}\,S(N+1)\,a,& 
\ds \hat V_{++}=\frac{\e^{-\beta\omega_0/2}}{\sqrt{Z_\beta}}\,C(N+1)^*,
\end{array}
\end{equation}
with $Z_\beta=\tr(\rho_{\beta,\cE})=1+\e^{-\beta\omega_0}$.

In particular $L_\beta$ is also a CP map and the above formula gives a Kraus representation for it. Moreover, since $\cL_\beta$ is trace preserving one easily gets that $\rhoinv^{1/2}$ is an invariant state of $L_\beta$ so that $\sr(L_\beta)=\|L_\beta\|=1$ where $\sr$ denotes the spectral radius. Note also that $\hat V_{-+}^*=\hat V_{+-}$ so that $L_\beta$ is self-adjoint on $\cJ_2$ if $C(N)$ is self-adjoint on $\cH_\cS$, which is the case when the detuning parameter $\eta$ vanishes (perfectly tuned cavity). In that case the analysis of the peripheral spectrum of $L_\beta$ is particularly simplified. In the general case, $L_\beta$ will more or less be a compact perturbation of a self-adjoint operator (see Lemma \ref{lem:Lbetadcompact}).


\subsection{Gauge invariance of $L_\beta$ and action on diagonal elements}\label{ssec:Lbetagaugeinv}

Let
\begin{equation*}\label{JdDef}
\cJ_2^{(d)}(\cH_\cS):= \{X \in \cJ_2(\cH_\cS)\,|\,\e^{-\i\theta N} X \e^{\i\theta N} =\e^{\i\theta d}X  \ \text{for all}\ \theta\in\R\},
\end{equation*}
(it is the set of bounded operators $X$ on $\cH_\cS=\ell^2(\N)$ which, in the eigenbasis of $H_\cS$, have the form $X=\sum_n x_n |n\ket\bra n+d|$ with $\sum_n|x_n|^2<\infty$), so that $\cJ_2(\cH_\cS)=\oplus_{d\in\Z}\,\cJ_2^{(d)}(\cH_\cS)$. It follows directly from the gauge invariance of the map $\cL_\beta$, see (\ref{eq:gauge}), that $L_\beta$ is also gauge invariant (this is also clear from its Kraus representation (\ref{eq:meanstate3})-(\ref{eq:vsigmahat})). It therefore leaves the subspaces $\cJ_2^{(d)}(\cH_\cS)$ invariant and hence admits a decomposition
\begin{equation*}\label{LDecomp}
L_\beta=\bigoplus_{d\in\Z} L_\beta^{(d)}.
\end{equation*}

In this section, we analyze the action of $L_\beta$ on diagonal elements of $\cJ_2$, i.e. the operator $L_\beta^{(0)}$. Denoting by $x_n$ the diagonal elements of $X\in \cJ_2^{(0)}$ we can identify $\cJ_2^{(0)}$ with $\ell^2(\N)$, and we immediately get from the Kraus representation (\ref{eq:meanstate3})-(\ref{eq:vsigmahat}) that
\begin{align*}
\label{eq:ldiag}
(L_\beta^{(0)}x)_n =  \frac{1}{Z_\beta}
&\left[\vphantom{\frac{\sin^2(\pi\sqrt{\xi(n+1)+\eta})}{\xi(n+1)+\eta}}
\left(\cos^2(\pi\sqrt{\xi n+\eta})
+\e^{-\beta\omega_0} \cos^2(\pi\sqrt{\xi(n+1)+\eta})\right)x_n \right. \\
&\left.+\frac{\sin^2(\pi\sqrt{\xi n+\eta})}{\xi n+\eta}
\left(\eta x_n+\e^{-\beta\omega_0/2}\xi nx_{n-1} \right)\right.\\
&\left.+\frac{\sin^2(\pi\sqrt{\xi(n+1)+\eta})}{\xi(n+1)+\eta}
\left(\e^{-\beta\omega_0}\eta x_n+\e^{-\beta\omega_0/2}\xi(n+1)x_{n+1}\right) \right].
\end{align*}

Following \cite{BP}, to rewrite this expression in a more convenient form we introduce the number operator
$$
(Nx)_n\equiv nx_n,
$$
as well as the twisted finite difference operators 
\begin{equation*}\label{eq:nablabeta}
(\nabla_\beta x)_n:= \left\{ \begin{array}{ll}x_0&\text{for}\ n=0,\\ x_n-\e^{-\beta\omega_0/2 }x_{n-1}& \text{for}\ n\ge1,
\end{array}
\right.
\qquad (\nabla_\beta^* x)_n:= x_n-\e^{-\beta\omega_0/2}x_{n+1},
\end{equation*}
on $\ell^2(\N)$. A simple algebra then leads to
\begin{equation}\label{Lbetanot}
L_\beta^{(0)}=\one-\nabla_\beta^* D(N) \nabla_\beta,
\end{equation}
where
\begin{equation}\label{Ddef}
D(N):= \frac{1}{Z_\beta}\,\sin^2(\pi\sqrt{\xi N+\eta})\, \frac{\xi N}{\xi N+\eta}.
\end{equation}
Note in particular that $L_\beta^{(0)}$ is self-adjoint.


\subsection{Spectral analysis of $L_\beta^{(0)}$}\label{ssec:Lbetanot}

The first result concerning the operator $L_\beta^{(0)}$ follows quite immediately from (\ref{Lbetanot})-(\ref{Ddef})
\begin{lemma}\label{lem:Lbetanotspectrum1} The operator $L_\beta^{(0)}$ satisfies
\begin{equation}\label{eq:Lbetanotbounds}
-1< -\frac{2\e^{-\beta\omega_0/2}}{1+\e^{-\beta\omega_0}} \leq L_\beta^{(0)} \leq 1,
\end{equation}
and $1$ is a simple eigenvalue, with eigenvector $\rhoinv^{1/2}$.
\end{lemma}

\proof 1. It follows directly  from (\ref{Ddef}) that $0\leq D(N)\leq \frac{1}{1+\e^{-\beta\omega_0}}$ so that
\begin{equation}\label{eq:Lbetanotbound1}
1 - \frac{1}{1+\e^{-\beta\omega_0}} \nabla_\beta^*\nabla_\beta  \leq L_\beta^{(0)} \leq 1.
\end{equation}
One then computes $\nabla_\beta^*\nabla_\beta= -\e^{-\beta\omega_0/2}\Delta +(1+\e^{-\beta\omega_0})$, where $(\Delta x)_n=x_{n+1}+x_{n-1}$ is the discrete Laplacian on $\ell^2(\N)$ with Dirichlet boundary condition, so that
$$
\nabla_\beta^*\nabla_\beta \leq 2\e^{-\beta\omega_0}+1.
$$
Combined with (\ref{eq:Lbetanotbound1}) we get (\ref{eq:Lbetanotbounds}).

2. $X$ is an eigenvector for the eigenvalue $1$ if and only if $\nabla_\beta^* D(N) \nabla_\beta X=0$. Since $\nabla_\beta^*$ is clearly injective we thus have $D(N)\nabla_\beta X=0$. It follows from (\ref{lock}) and (\ref{Ddef}) that $D(n)=0$ iff $n$ is a Rabi resonance. Since we are in a non-resonant situation $D(N)$ is injective as well. We end up with
$$
\nabla_\beta X=0 \quad \Longleftrightarrow X=C \e^{-\beta\omega_0N/2} =C\rhoinv^{1/2},
$$
i.e. $1$ is a simple eigenvalue with $\rhoinv^{1/2}$ as eigenvector.
\qed

The above lemma will be sufficient to prove that $L_\beta$ is mixing, i.e. Theorem \ref{thm:mixingLbeta}. By mimicking the proof of Lemma \ref{lem:cLbetanotspectrum} we can actually also prove the following
\begin{prop}\label{prop:Lbetanotspectrum2} $1$ is in the essential spectrum of $L_\beta^{(0)}$.
\end{prop}
Such a result would be usefull to prove that $L_\beta$ is slowly mixing. We shall however directly prove that $\cL_\beta$ is slowly mixing without using a similar property for $L_\beta$.

\begin{remark}\label{rem:noacspectrum} The nature of the spectrum of $L_\beta^{(0)}$ plays no role in our mixing results. However it follows almost immediately from the same argument which leads to Proposition \ref{prop:Lbetanotspectrum2} that this spectrum is purely singular. Let 
$$
L_{\beta,0}^{(0)}:= \one-\nabla_\beta^* D_0(N) \nabla_\beta,
$$
where $D_0(N)$ is defined as in (\ref{def:Dnot}). Then $L_{\beta}^{(0)}- L_{\beta,0}^{(0)}$ is a trace class operator. Decomposing $\ell^2(\N)$ as $\ell^2(\N)=\oplus_k \ell^2(\{m_k,\ldots,m_{k+1}-1\})$, where the $m_k's$ are defined in (\ref{eq:quasirabidecay}), it is easy to see, using $D_0(m_k)=0$ for any $k$, that $L_{\beta,0}^{(0)}$ leaves each $\ell^2(\{m_k,\ldots,m_{k+1}-1\}$ invariant. Since all these subspaces are finite dimensional the spectrum of $L_{\beta,0}^{(0)}$ is actually pure point. By trace class perturbation, see e.g. Sect. X.4. in \cite{K} , the spectrum of $L_\beta^{(0)}$ is indeed purely singular. We however do not know the precise nature of this spectrum. 
\end{remark}


\subsection{Spectral analysis of $L_\beta^{(d)}$}\label{ssec:Lbetad}

We shall further use the spectral results about the operator $L_\beta^{(0)}$ to get information about the operator $L_\beta^{(d)}$ for arbitrary $d\in\Z$. The goal of this section is to prove

\begin{prop}\label{prop:Lbetadspectrum} For any $d\neq 0$, $\sp(L_\beta^{(d)}) \cap S^1 =\{1\}$ and $1$ is not an eigenvalue.
\end{prop}

Denoting by $x_n$ the coefficients of $X\in \cJ_2^{(d)}$, i.e. $X=\sum_n x_n\, |n\ket\bra n+d|$ (the sum starts at $\max\{0,-d\}$), we can identify $\cJ_2^{(d)}$ with $\ell^2(\N)$, and we immediately get from the Kraus representation (\ref{eq:meanstate3})-(\ref{eq:vsigmahat}) that
\begin{eqnarray*}
(L_\beta^{(d)}x)_n & = &  \frac{1}{Z_\beta} \left[\left(\overline{C(n)}C(n+d) +\e^{-\beta\omega_0} C(n+1)\overline{C(n+1+d)}\right)x_n \right. \nonumber \\
 & & \qquad +\e^{-\beta\omega_0/2} \sqrt{n(n+d)}\,\overline{S(n)}S(n+d) x_{n-1} \nonumber \\
 & & \qquad +\e^{-\beta\omega_0/2}\sqrt{(n+1)(n+d+1)}\,\overline{S(n+1)}S(n+d+1)x_{n+1} \Big].
\end{eqnarray*}
where $C(n)$ and $S(n)$ are defined in (\ref{def:CSfunctions}). A simple algebra leads to
\begin{eqnarray}\label{eq:loffdiag}
L_\beta^{(d)} & = & \frac{1}{Z_\beta} \Big[ C(N)^*C(N+d)+\e^{-\beta\omega_0} C(N+1)C(N+d+1)^* \\
 & & \qquad  +\tilde S(N)^*\tilde S(N+d) (\one-\nabla_\beta)+ (\one-\nabla_\beta^*)\tilde S(N)^*\tilde S(N+d)\Big], \nonumber
\end{eqnarray}
where $N$,$\nabla_\beta$ and $\nabla_\beta^*$ are as in Section \ref{ssec:Lbetagaugeinv} and $\tilde S(N)=\sqrt{N}S(N)$.

Via this identification, we can consider that both $L_\beta^{(d)}$ and $L_\beta^{(0)}$ act on $\ell^2(\N)$. An easy calculation shows that $C(n+d)-C(n)=O\left(n^{-1/2}\right)$ and $\tilde S(n+d)-\tilde S(n)=O\left(n^{-1/2}\right)$. Since the operators $C(N)$, $\tilde S(N)$ and $\nabla_\beta$ are bounded we get from (\ref{Lbetanot}), (\ref{eq:loffdiag}) and the fact $C(N)^*C(N)+\tilde S(N)^*\tilde S(N) = \one$ the following
\begin{lemma}\label{lem:Lbetadcompact} For any $d\in\Z$, the operator $L_\beta^{(d)}-L_\beta^{(0)}$ is compact. 
\end{lemma}

As a consequence $\sp_\ess(L_\beta^{(d)})=\sp_\ess(L_\beta^{(0)})$, and in particular
$$
1\in \sp_\ess(L_\beta^{(d)})\subset \left[-\frac{2\e^{-\beta\omega_0/2}}{1+\e^{-\beta\omega_0}} ,1\right],
$$
so that $\sp(L_\beta^{(d)})\cap S^1 =\{1\}\cup \left(\sp_{\rm disc}(L_\beta^{(d)})\cap S^1  \right)$. The following Lemma shows that $L_\beta^{(d)}$ does not have eigenvalues on $S^1$ for $d\neq 0$. This completes the proof of Proposition \ref{prop:Lbetadspectrum}.

\begin{lemma}\label{lem:peripheralev} The only peripheral eigenvalue of $L_\beta$ is $1$ and it is simple, with invariant vector $\rhoinv^{1/2}\in \cJ_2^{(0)}$. In particular $\sp_{\rm disc} (L_\beta^{(d)})\cap S^1 = \emptyset$ for any $d\neq 0$.
\end{lemma}

As we mentioned in Section \ref{ssec:Lbetadef}, $L_\beta$ is a completely positive operator with $\sr(L_\beta)=\|L_\beta\|=1$ so we can apply Theorem \ref{thm:schrader}.

\noindent {\bf Proof of Lemma \ref{lem:peripheralev}}\ \ Let $\theta\in\R$ and $X\in \cJ_2$ such that $L_\beta(X)=\e^{i\theta} X$. According to the decomposition (\ref{LDecomp}), it suffices to consider $X\in \cJ_2^{(d)}$. By Lemma \ref{lem:Lbetanotspectrum1}, we only need to consider $d\neq 0$.  

Note that $X^*\in \cJ_2^{(-d)}$ satisfies then $L_\beta(X^*)=\e^{-i\theta}X^*$ so that, by Theorem \ref{thm:schrader}, both $Y=\sqrt{X^*X}\in \cJ_2^{(0)}$ and $Z=\sqrt{XX^*}\in\cJ_2{(0)}$ are invariant vectors. It follows from Lemma \ref{lem:Lbetanotspectrum1} that $Y$ and $Z$ are proportional to $\rhoinv^{1/2}>0$. Since either $Y$ (if $d>0$) or $Z$ (if $d<0$) has a non-trivial kernel, this proves that either $Y$ or $Z$ is zero and hence $X=0$.
\qed

\begin{remark}\label{rem:Lbeta*invstate} The same reasoning applies to the operator $L_\beta^*$ and, since $L_\beta^{(0)}$ is selfadjoint, shows that 
$$
\ker(L_\beta^*-\one)=\ker\left(L_\beta^{(0)}-\one\right)=\C\rhoinv^{1/2}.
$$
\end{remark}


\subsection{Mixing properties of $L_\beta$}\label{ssec:Lbetamixing}

The purpose of this section is to prove the following

\begin{theorem}\label{thm:mixingLbeta} The iterates of $L_\beta$ converge strongly to $|\rhoinv^{1/2}\ket\bra \rhoinv^{1/2}|$, i.e.
\begin{equation}\label{eq:Lbetamixing}
\lim_{n\to\infty} L_\beta^n(X) = \tr\left(\rhoinv^{1/2}X \right) \rhoinv^{1/2},\qquad \forall X\in \cJ_2.
\end{equation}
\end{theorem}

Our main tool is Theorem \ref{thm:bgm1}. Although, for any $d\in\Z$, $\sp\left(L_\beta^{(d)}\right)\subset \D\cup \{1\}$ we do not have such an inclusion for $L_\beta$: $\ds \sp(L_\beta)= \overline{\cup_{d\in\Z}  \sp\left(L_\beta^{(d)}\right)}$ and we may have eigenvalues of $L_\beta^{(d)}$ which accumulate toward the unit circle when $d$ becomes large. We shall bypass this issue using the following approximation argument

\begin{lemma}\label{lem:approximation} For any $X\in \cJ_2(\cH_\cS)$, there exists $(X_k)_k$ such that 
$$
X_k\in \cJ_2^{(\leq k)}:= \bigoplus_{|d|\leq k}\,\cJ_2^{(d)}(\cH_\cS)
$$
and $\ds \lim_{k\to\infty}X_k=X$ in $\cJ_2(\cH_\cS)$.
\end{lemma}

\proof If $X=\sum_{n,m} x_{nm} |n\ket\bra m| \in \cJ_2$, it suffices to take $X_k:= \sum_{|n-m|\leq k} x_{nm} |n\ket\bra m|$.
\qed

\noindent {\bf Proof of Theorem \ref{thm:mixingLbeta}. \ \ } First note that since $L_\beta$ is a contraction on the Hilbert space $\cJ_2$, the von Neumann mean ergodic theorem asserts that
$$
\slim_{N\to\infty} \frac{1}{N} \sum_{n=1}^N L_\beta^n = P
$$
where $P$ is the projection onto $\ker(L_\beta-\one)$ along $\overline{\ran(L_\beta-\one)}=\left( \ker(L_\beta^*-\one) \right)^\perp$. By Lemma \ref{lem:Lbetanotspectrum1} and Remark \ref{rem:Lbeta*invstate} we have $\ker(L_\beta-\one)=\ker(L_\beta^*-\one)=\C \rhoinv^{1/2}$ so that 
$P=|\rhoinv^{1/2}\ket\bra \rhoinv^{1/2}|$.

We will prove (\ref{eq:Lbetamixing}) for $X\in \bigoplus_{|d|\leq k}\,\cJ_2^{(d)}(\cH_\cS)$ where $k\in\N$ is fixed. The result then follows from Lemma \ref{lem:approximation} since the left hand side of (\ref{eq:Lbetamixing}) is continuous in $X$ uniformly in $n$ while the right-hand side is continuous in $X$.

For any given $k$, $\cJ_2^{(\leq k)}$ is a closed invariant subspace for $L_\beta$ and
$$
\sp(L_\beta \lceil_{\cJ_2^{(\leq k)}})= \cup_{|d|\leq k} \sp\left(L_\beta^{(d)}\right),
$$
so that, using Lemma \ref{lem:Lbetanotspectrum1} and Proposition \ref{prop:Lbetadspectrum}, we have 
$$
\sp\left(L_\beta\lceil_{\cJ_2^{(\leq k)}}\right) \subset \D\cup \{1\}.
$$
We can therefore apply Theorem \ref{thm:bgm1} which proves (\ref{eq:Lbetamixing}) if $X\in \cJ_2^{(\leq k)}$.
\qed

\begin{remark} One can also prove that the mixing is slow, i.e. $L_\beta$ satisfies (ASC). Indeed, since $1$ is a simple eigenvalue and belongs to the essential spectrum of $L_\beta$ one has
$$
\sr\left( L_\beta \lceil_{\overline{\ran(L_\beta-\one)}}\right)=1
$$
so that, by Theorem \ref{thm:bgm2}, $L_\beta$ satisfies (ASC).
\end{remark}


\subsection{Proof of Theorem \ref{thm:strongmixing}}\label{sec:proofthm1}

Since $\cL_\beta$ is a contraction and finite rank operators are dense in $\cJ_1(\cH_\cS)$ it suffices to prove the result for initial states $\rho$ which are finite rank operators. Then, because the non-interacting evolution $\cU$, see (\ref{def:freeevol}), preserves the trace norm, using (\ref{eq:dynamicsplitting}) and the fact that $\rhoinv$ is $\cU$-invariant, it suffices to prove that
\begin{equation}\label{eq:Ltildemixing}
\lim_{n\to \infty}  \left\|\tilde \cL_\beta^n(\rho)-\rhoinv\right\|_1=0.
\end{equation}
Moreover, because $\tilde \cL_\beta$ is a completely trace preserving map and $\rho$ is a state, one has for any $n$
$$
\left\| \tilde \cL_\beta^n(\rho) \right\|_1 = \tr \left(\tilde \cL_\beta^n(\rho)\right) = 1 = \| \rhoinv\|_1,
$$
and in particular $\ds \lim_{n\to \infty} \left\|  \tilde \cL_\beta^n(\rho)\right\|_1 = \left\|  \rhoinv\right\|_1$. To prove (\ref{eq:Ltildemixing}) it therefore suffices to prove that, see \cite{S},
\begin{equation}\label{eq:Ltildewmixing}
\lim_{n\to \infty}  \tr\left[\tilde \cL_\beta^n(\rho) A\right]=\tr\left[\rhoinv A\right], \qquad \forall A\in\cB(\cH_\cS).
\end{equation}

Let therefore $\rho$ be an initial state with finite rank and $A\in \cB(\cH_\cS)$. We have
\begin{eqnarray*}
\tr\left[\tilde \cL_\beta^n(\rho) \times A \right] & = & \tr\left[ \rhoinv^{-1/4}\, \rho \, \rhoinv^{-1/4} \times \Phi\circ (\tilde \cL_\beta^*)^n(A) \right]\\
 & = & \tr\left[ \rhoinv^{-1/4}\, \rho \, \rhoinv^{-1/4} \times L_\beta^n(\Phi(A)) \right],
\end{eqnarray*}
where we used the cyclicity of the trace in the second line ($\rhoinv^{-1/4}\rho \rhoinv^{-1/4}$ is a well defined trace class operator since $\rho$ has finite rank and $\rhoinv>0$). By Theorem \ref{thm:mixingLbeta} we thus have
\begin{eqnarray*}
\lim_{n\to \infty} \tr\left[\tilde \cL_\beta^n(\rho) \times A \right] & = & \tr\left[ \rhoinv^{-1/4}\, \rho\,  \rhoinv^{-1/4} \times  \rhoinv^{1/2}\right] \times \tr\left[ \rhoinv^{1/2}  \Phi(A)\right] \ = \ \tr\left[\rhoinv  A\right],
\end{eqnarray*}
which proves (\ref{eq:Ltildewmixing}).

\bigskip

Decomposing an element $X\in \cJ_1(\cH_\cS)$ as $X=X_{r,+}-X_{r,-}+\i (X_{i,+}-X_{i,-})$, with $X_{r/i,\pm}$ positive, one then actually gets

\begin{corollary}\label{coro:strongcvg} Under the hypotheses of Theorem \ref{thm:strongmixing}, $\ds \slim_{n\to\infty} \cL_\beta^n =\cL_\beta^\infty$ on $\cJ_1(\cH_\cS)$ where $\cL_\beta^\infty(X):= \tr(X) \rhoinv$. 
\end{corollary}


\section{Proof of Theorem \ref{thm:slowmixing}}\label{sec:proof2}

Besides the notion of arbitrarily slow convergence, the authors of \cite{BGM} also introduce the notion of quick uniform convergence (QUC) if there exists $C>0$ and $\alpha \in \,]0,1[$ such that $\|T^n-T_\infty\|\leq C\alpha^n$ for all $n$. Note that the latter implies in particular exponential mixing. The main ingredient in the proof of Theorem \ref{thm:slowmixing} is the following result due to Badea, Grivaux and M\"uller \cite{BGM}.

\begin{theorem}\label{thm:bgm2} Let $Z$ be a Banach space which contains no isomorphic copy of $c_0$\footnote{$c_0$ denotes the Banach space of complex sequences which converge to $0$ (endowed with the $\ell^\infty$ norm).} and $T\in\cB(Z)$ such that the sequence of iterates $T^n$ is strongly convergent to $T^\infty\in\cB(Z)$. Then the following dichotomy holds: $T$ satisfies either (QUC) or (ASC). Moreover (QUC) holds if and only if $\sr\left(T\lceil_{\overline{\ran(T-\one)}}\right)<1$. 
\end{theorem}

\begin{remark} The necessary and sufficient condition for (QUC) is not stated in this form in \cite{BGM} but it appears explicitly in the proof of their theorem. 
\end{remark}

We shall apply Theorem \ref{thm:bgm2} to $T=\cL_\beta$ acting on $Z=\cJ_1(\cH_\cS)$. Note that $\cJ_1(\cH_\cS)$ indeed contains no isomorphic copy of $c_0$. We then have to prove that the initial vector in $\cJ_1(\cH_\cS)$ such that (\ref{eq:ascw}) holds can be chosen as a state.


\subsection{Block structure and essential spectral radius of $\cL_\beta$}\label{ssec:essentialradius}

Since the operator $\cL_\beta$ is gauge invariant, see (\ref{eq:gauge}), it can be decomposed in a similar way as $L_\beta$. If
$$
\cJ_1^{(d)}(\cH_\cS):= \{X \in \cJ_1(\cH_\cS)\,|\,\e^{-\i\theta N} X \e^{\i\theta N} =\e^{\i\theta d}X  \ \text{for all}\ \theta\in\R\},
$$
$\cL_\beta$ leaves $\cJ_1^{(d)}(\cH_\cS)$ invariant and thus admits a decomposition
\begin{equation*}\label{cLDecomp}
\cL_\beta=\bigoplus_{d\in\Z} \cL_\beta^{(d)}.
\end{equation*}

In view of Theorem \ref{thm:bgm2} we are interested in the spectral radii of the $\cL_\beta^{(d)}$ restricted to $\overline{\ran(\cL_\beta^{(d)}-\one)}$. In this section we prove the following

\begin{prop}\label{prop:essentialsr} For any $d\in\Z$, $\e^{i\omega \tau d}\in \sp_\ess\left(\cL_\beta^{(d)} \right)$. As a consequence, for any $d\in \Z$,
$$
\sr\left(\cL_\beta^{(d)} \lceil_{\overline{\ran(\cL_\beta^{(d)}-\one)}}\right)=1.
$$
\end{prop}

\begin{remark} For any $d\in \Z$, $\e^{i\omega \tau d}\in \sp_\ess\left(\cL_\beta\right)$. In particular, as mentionend in Remark \ref{rem:uncoupledspectrum}, the spectrum of the uncoupled reduced dynamics operator survives when one turns on the interaction. 
\end{remark}

As in Section \ref{sec:proof1} we shall first obtain information on $\cL_\beta^{(0)}$ and then derive information for $\cL_\beta^{(d)}$, $d\neq 0$.

\begin{lemma}\label{lem:cLbetanotspectrum} $1\in \sp_\ess\left(\cL_\beta^{(0)} \right)$. 
\end{lemma}

From its Kraus representation (\ref{eq:meanstate})-(\ref{eq:vsigma}), and up to identifying $\cJ_1^{(0)}(\cH_\cS)$ with $\ell^1(\N)$, one gets an expression similar to (\ref{Lbetanot}) for $\cL_\beta^{(0)}$: 
$$
\cL_\beta^{(0)} = \one - \nabla_0^* D(N) \nabla_{2\beta}.
$$
\begin{remark} From the above formula one may, at least formally, write $\cL_\beta^{(0)}$ as 
$$
\cL_\beta^{(0)}=\e^{\beta\omega_0 N/2} L_\beta^{(0)} \e^{\beta \omega_0 N/2}.
$$
This explains the origin of the embedding $\Phi$ used in Section \ref{sec:proof1}. 
\end{remark}

We shall prove that $\cL_\beta^{(0)}$ is actually a compact perturbation of an operator which has $1$ as an infinitely degenerate eigenvalue. For that purpose we recall the notion of Rabi quasi-resonance introduced in \cite{BP} and already mentioned in Section \ref{ssec:ergodicity}.

\begin{definition}\label{def:quasirabi} We say that $m\in\N^\ast$ is a Rabi quasi-resonance if it satisfies $D(m)<D(m\pm1)$.
\end{definition}
Let $(m_k)_{k\in\N^*}$ be the strictly increasing sequence of quasi-resonances. It is straightforward to show that 
\begin{equation}\label{eq:quasirabidecay}
 D(m_k)=O(k^{-2}) \quad {\rm as} \quad k\to\infty. 
\end{equation}

\noindent {\bf Proof of Lemma \ref{lem:cLbetanotspectrum}.} \ \ Let
\begin{equation}\label{def:Dnot}
D_0(n):=\left\{\begin{array}{ll}
0 & \text{if}\ n\in\{m_1,m_2,\ldots\},\\
D(n)&\text{otherwise},
\end{array}
\right.
\end{equation}
and
\begin{equation*}\label{def:Lbetanotnot}
\cL_{\beta,0}^{(0)}:= \one-\nabla_0^* D_0(N) \nabla_{2\beta}.
\end{equation*}
It immediately follows from (\ref{eq:quasirabidecay}) that the operator $D(N)-D_0(N)$ is compact. Since $\nabla_{2\beta}$ and $\nabla_0^*$ are bounded we get that 
\begin{equation*}
\cT:= \cL_{\beta}^{(0)}- \cL_{\beta,0}^{(0)},
\label{essdec}
\end{equation*}
is a compact operator as well. 

A similar argument to the one of Lemma \ref{lem:Lbetanotspectrum1} (see also \cite{BP}, Section 4.5.3) shows that $1$ is an infinitely degenerate eigenvalue of $\cL_{\beta,0}^{(0)}$ with corresponding (normalized) eigenvectors
$$
\rho_k:= \frac{\e^{-\beta\omega_0 N}\widetilde P_k}{\tr\left(\e^{-\beta\omega_0 N}\widetilde P_k\right)},
$$
where $\widetilde P_k$ denotes the orthogonal projection onto $\ell^2(\{0,\ldots,m_k-1\})$, and with $m_0=0$. Indeed, $\rho$ is an invariant vector iff $D_0(N)\nabla_{2\beta}\, \rho=0$, and because $D_0$ vanishes at the $m_k's$ the eigenvalue equation splits into an infinite number of finite dimensional systems
$$
\rho_n=\e^{-\beta\omega_0}\rho_{n-1}, \quad n\in{m_{k-1}+1,\ldots, m_k-1}. 
$$
In particular, $1$ is in the essential spectrum of $\cL_{\beta,0}^{(0)}$ and hence of $\cL_\beta^{(0)}$.
\qed

The presence of these quasi-resonances imply that the ``quasi Rabi sectors'' $\ell^1(\{m_k,\ldots,m_{k+1}-1\})$ are very weakly coupled for large $k$. 
The vectors $\rho_k$ are the metastable (or almost invariant) states we already mentioned and which are at the origin of the slow relaxation.

\noindent {\bf Proof of Proposition \ref{prop:essentialsr}.}\ \  Denoting by $x_n$ the coefficients of $X\in \cJ_1^{(d)}$, i.e. $X=\sum_n x_n |n\ket\bra n+d|$, we identify $\cJ_1^{(d)}$ with $\ell^1(\N)$. We can then proceed as in Section \ref{ssec:Lbetad} to prove that via this identification the operator $\cL_\beta^{(d)}-\e^{i\omega\tau d}\cL_\beta^{(0)}$ is compact and the first part then follows from Lemma \ref{lem:cLbetanotspectrum}. We leave the details to the reader.

Since $1$ is a simple eigenvalue of $\cL_\beta$ (see Theorem \ref{thm:BP}), and actually of $\cL_\beta^{(0)}$, we get that for any $d\in \Z$ 
$$
\e^{i\omega \tau d}\in \sp\left(\cL_\beta^{(d)} \lceil_{\overline{\ran(\cL_\beta^{(d)}-\one)}}\right)
$$
(it is actually trivial when $\e^{i\omega \tau d}\neq 1$).
\qed


\subsection{Proof of Theorem \ref{thm:slowmixing}}\label{ssec:slowmixingproof}

Combining Corollary \ref{coro:strongcvg}, Proposition \ref{prop:essentialsr} together with Theorem \ref{thm:bgm2} we immediately get the following

\begin{prop}\label{prop:ascw} For any $d\in\Z$ the operator $\cL_\beta^{(d)}$ satisfies (ASC). 
\end{prop}

\noindent {\bf Proof of Theorem \ref{thm:slowmixing}.}\ \ It follows from the above proposition that $\cL_\beta$ satisfies (ASC). It remains to show that in (\ref{eq:slowmixing}) we can indeed chose $\rho$ to be a state.

Let $(\epsilon_n)_n$ be a sequence of positive numbers and let $d\neq 0$. Since $\cL_\beta^{(d)}$ satisfies (ASC) there exist $X\in \cJ_1^{(d)}$, $A\in \cB(\cH_\cS)$ and $n_0\in\N$ such that
$$
|\tr(\cL_\beta^n(X) \, A) -\tr(\cL_\beta^\infty(X) \, A) | = |\tr(\cL_\beta^n(X) \, A)|\geq \epsilon_n, \qquad \forall n\geq n_0.
$$
(Note that when $d\neq 0$ one has $\tr(X)=0$ so that $\cL_\beta^\infty(X)=0$.) One can actually assume that 
$$
A\in \cB^{(-d)}:=\{A \in \cB(\cH_\cS)\,|\,\e^{-\i\theta N} A \, \e^{\i\theta N} =\e^{-\i\theta d}A  \ \text{for all}\ \theta\in\R\}.
$$
Indeed, $\cL_\beta$ leaves $\cJ_1^{(d)}$ invariant and if $X\in \cJ_1^{(d)}$ and $A\in \cB^{(k)}$ then $\tr(XA)=0$ if $k\neq -d$.

The operator $X+X^*+|X|+|X^*|$ is then positive. Moreover $X^*\in \cJ_1^{(-d)}$ and $|X|+|X^*|\in \cJ_1^{(0)}$ so that
\begin{equation}\label{eq:slowmixingproof}
\left|\tr \left( \cL_\beta^n(X+X^*+|X|+|X^*|) \, A \right)   \right| = |\tr (\cL_\beta^n(X) \, A)|\geq \epsilon_n, \qquad \forall n\geq n_0.
\end{equation}
It thus remains to take $\rho= \frac{X+X^*+|X|+|X^*|}{\tr(X+X^*+|X|+|X^*|)}$ (recall that $\tr(\rhoinv \, A)=0$ for $A\in \cB^{(-d)}$).
\qed

As mentioned in Remark \ref{rem:slowdecoherence}, the above proof shows that the decoherence in the energy eigenbasis of the cavity field is arbitrarily slow too: inequality (\ref{eq:slowmixingproof}) is due to the off-diagonal part of $\cL_\beta^n(X+X^*+|X|+|X^*|)$. Actually, our proof of Theorem \ref{thm:slowmixing} could give the impression that the slowness of the mixing is only due to slow decoherence (we started from $X\in \cJ_1^{(d)}$, $d\neq0$). The following Proposition shows that this is not the case and one can also have slow mixing starting from an initial state $\rho\in \cJ_1^{(0)}$.

\begin{prop}\label{prop:slowmixingdiagonal} Suppose the system is non-resonant and $\beta>0$. Then for any sequence $(\epsilon_n)_n$ of positive numbers such that $\lim \epsilon_n=0$, there exist an initial state $\rho\in \cJ_1^{(0)}$, an observable $A$, and $C>0$ such that (\ref{eq:slowmixing}) holds up to extracting a subsequence.
\end{prop}

\proof Given a sequence  $(\epsilon_n)_n$, from Proposition \ref{prop:ascw} there exist $X\in \cJ_1^{(0)}$, $A\in \cB(\cH)$ and $n_0$ such that
\begin{equation}\label{eq:slowdiagonal}
|\tr (\cL_\beta^n(X)\, A)-\tr(X) \tr(\rhoinv\, A) |\geq \epsilon_n, \qquad \forall n\geq n_0.
\end{equation}
Writing $X=X_{r,+}-X_{r,-}+\i(X_{i,+}-X_{i,-})$, with $X_{r/i,\pm}$ positive, we have for all $n\geq n_0$
\begin{eqnarray*}
\epsilon_n & \leq & |\tr (\cL_\beta^n(X_{r,+})\, A)-\tr(X_{r,+}) \tr(\rhoinv\, A) | + |\tr (\cL_\beta^n(X_{r,-})\, A)-\tr(X_{r,-}) \tr(\rhoinv\, A) | \\
 & & + |\tr (\cL_\beta^n(X_{i,+})\, A)-\tr(X_{i,+}) \tr(\rhoinv\, A) | + |\tr (\cL_\beta^n(X_{i,-})\, A)-\tr(X_{i,-}) \tr(\rhoinv\, A) |.
\end{eqnarray*}
Up to extracting a subsequence (\ref{eq:slowdiagonal}) therefore holds for at least one of the $X_{r/i,\pm}$. It suffices to take $\ds \rho = \frac{X_{r/i,\pm}}{\tr(X_{r/i,\pm})}$ (the trace can not be $0$ since $X_{r/i,\pm}\geq 0$ and (\ref{eq:slowdiagonal}) holds).
\qed

\begin{remark} That one has to extract a subsequence is certainly an artefact of our proof and Proposition \ref{prop:slowmixingdiagonal} certainly holds without such an extraction (recall that $\cL_\beta$ is a contraction so that one can not expect that the convergence would be ``fast'' along another subsequence). However, even if it holds only up to a subsequence, Proposition \ref{prop:slowmixingdiagonal} shows that there is also no lower bound on the speed of convergence in $\cJ_1^{(0)}$. 
\end{remark}

\begin{remark} As mentioned in Remark \ref{rem:noacspectrum} the spectrum of the operator $L_\beta^{(0)}$ is purely singular. A further analysis of the latter would be important to investigate an upper bound on the convergence speed, e.g. the presence of point spectrum would lead to an exponential upper bound on the convergence speed for $L_\beta$.
\end{remark}


\end{document}